\documentstyle[12pt,psfig]{article}   %%LaTeX 2.09

\newcommand{\beq}{\begin{equation}}
\newcommand{\eeq}{\end{equation}}
\newcommand{\bea}{\begin{eqnarray}}
\newcommand{\eea}{\end{eqnarray}}
\def\lappeq{\mathrel{\rlap {\raise.5ex\hbox{$<$}}
            {\lower.5ex\hbox{$\sim$}}}}
\def\pl#1#2#3{{\it Phys. Lett. }{\bf B#1} (19#2) #3}
\def\zp#1#2#3{{\it Z. Phys. }{\bf C#1} (19#2) #3}
\def\prl#1#2#3{{\it Phys. Rev. Lett. }{\bf #1} (19#2) #3}
\def\pr#1#2#3{{\it Phys. Rev. }{\bf D#1} (19#2) #3}
\def\np#1#2#3{{\it Nucl. Phys. }{\bf B#1} (19#2) #3}
\def\app#1#2#3{{\it Acta Phys. Polon. }{\bf #1} (19#2) #3}

\begin{document}              

%\begin{titlepage}
\begin{flushright}
{\small
INFN-ISS/98-1\\
ROME1-1194-98\\
TUM-HEP-307/98\\
January 1998
}
\end{flushright}

\begin{center}
\vspace{0.8cm}
\renewcommand{\thefootnote}{\fnsymbol{footnote}}
{\large\bf Factorization, charming penguins, and all
that}\footnote{Talk given by M.C. at Beauty '97, 5th Intl. Workshop on
$B$ Physics at Hadron Machines, UCLA, Los Angeles, CA, USA,
October 13--17, 1997.}\\
\vspace{0.8cm}
{\bf    M.~Ciuchini$^a$,
        R.~Contino$^b$, E.~Franco$^b$, G.~Martinelli$^b$, and
        L.~Silvestrini$^c$
}
\setcounter{footnote}{0}
\vspace{.4cm}

{\it
        $^a$ INFN, Sez. Sanit\`a, V.le Regina Elena, 299, I-00186, Rome,
        Italy\\
\vspace{2mm}
        $^b$ Dip. di Fisica, Universit\`a ``La Sapienza'' and
        INFN, Sez. di Roma, P.le~A.~Moro, 3, I-00185, Rome, Italy \\
\vspace{2mm}
        $^c$ Physik Dept., Technische Universit\"at M\"unchen,
        D-85748 Garching, Germany\\
}
\vspace{1cm}
{\large\bf Abstract}
\end{center}
We discuss few selected topics related to the calculation of
hadronic amplitudes relevant for two-body non-leptonic $B$ decays.
%\end{titlepage}

\vspace{0.4cm}

\subsection*{Introduction}
It is likely
that most of the future results in $B$ phenomenology, including
the study of CP violation, will come from the measurements of two-body
non-leptonic decays.
The major theoretical problem in predicting the rates of these decays
is the evaluation of hadronic amplitudes. 
In general the solution to this problem is unknown as it
contains all the difficulties of low-energy strong interactions and
hadronization. In QCD, the best one can do is using the operator product
expansion to separate the short- and long-distance scales. In the
resulting effective Hamiltonian, the effect of short-distance physics can
be computed perturbatively and it is described by the notorious Wilson
coefficients. Long-distance, non-perturbative physics is contained in the
hadronic matrix elements of a set of local operators. In particular, for
$B$ decays into two mesons, one needs to compute matrix elements of
dimension-six four-fermion operators (we neglect in the following
magnetic dipole transitions),
\beq
\langle M_1 M_2\vert Q_i\vert B\rangle=
\langle M_1 M_2\vert \bar b\Gamma_i q_1~
\bar q_2\Gamma_i^\prime q_3\vert B\rangle~,
\qquad i=1,\dots,10~,
\eeq
where $b$, $q_j$ are the appropriate quark fields and
$\Gamma_i$, $\Gamma^\prime_i$ are various combinations of Dirac and colour
matrices. Unlike $K$ physics, neither analytic computation
techniques nor systematic expansions are known for these matrix elements.
Numerical approaches to QCD are severely limited by the present
computing power, since a small lattice spacing is required to simulate
the heavy $B$ field.

These problems have not prevented theorists from making predictions of
interesting quantities, such as BRs, asymmetries, etc.
On the one hand, observables free from hadronic uncertainties have
been identified, the best (and apparently unique) example being the CP
asymmetry in $B_d\to J/\Psi K_S$. On the other hand, when the evaluation
of hadronic matrix elements could not be avoided, as for the BRs, various
theoretical approaches, such as flavour symmetry, factorization and form-factor
models, have been developed and used.

Although these approaches have been successful in some applications,
their theoretical soundness is questionable. From a phenomenological
point of view, this means that the theoretical error affecting the
predicted amplitudes can only be guessed.

\subsection*{Factorization}

\begin{figure}[t]
\centerline{
\psfig{figure=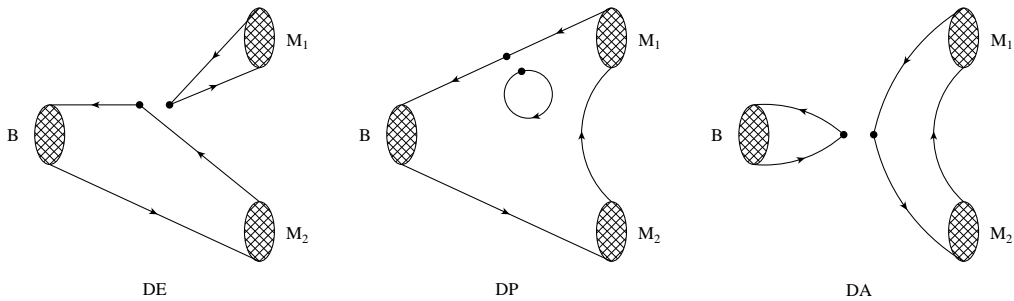,width=.9\textwidth}
}
\caption{\sf 
Diagrams representing the Wick contraction of a four-fermion operator between
a $B$ and two mesons $M_1$ and $M_2$. The operator insertion is represented
by the double black dots. Disconnected emission ($DE$), penguin ($DP$)
and annihilation ($DA$) diagrams are shown.
}
\label{fig:diag}
\end{figure}

To be concrete, let us consider the popular approach based on the
factorization hypothesis applied to an emission-dominated decay, say
$B^+\to D^+\pi^0$. Using Fierz and colour rearrangement, one is left only
with the matrix element
\beq
\langle D^+\pi^0\vert \bar b \gamma_\mu (1-\gamma_5) u
~\bar c \gamma^\mu (1-\gamma_5) d \vert B^+\rangle~,
\eeq
where the quark fields are contracted according to the disconnected
emission diagram $DE$ in fig.~\ref{fig:diag}.
The factorization hypothesis states that
\beq
\langle D^+\pi^0\vert \bar b \gamma_\mu (1-\gamma_5) u
~\bar c \gamma^\mu (1-\gamma_5) d \vert B^+\rangle=
\langle \pi^0\vert \bar b \gamma_\mu (1-\gamma_5) u
\vert B^+\rangle\langle D^+\vert
\bar c \gamma^\mu (1-\gamma_5) d \vert 0\rangle~,
\eeq
namely the four-fermion operator matrix element is given by the product of
the matrix elements of two currents. Pictorially this means that the
two disconnected branches of the emission diagram do not interact.
The heuristic physical argument is appealing~\cite{bj}: {\it if} a large
energy is transferred to the emitted meson, as in the case at hand, the
emitted quarks have not ``enough'' time to interact before going far from
the interaction point and hadronizing. Starting from this picture, Dugan and
Grinstein introduced the large energy effective theory (LEET)~\cite{leet},
in which factorization holds at the lowest order of a systematic expansion
in powers of $\Lambda_{QCD}/E_{emit}$.
Besides the objections that can be raised even on the lowest-order
result~\cite{ugo}, higher order corrections in LEET, or equivalently
corrections to factorization, are not known. This is already a crucial
point: even accepting that
factorization holds in some effective expansion of QCD, there is {\it no}
indication that it is an accurate approximation, and corrections at the
level of $10$--$20\%$, or even more, are not surprising at all.

\subsection*{Charming penguins}

It is easy to find decays where a moderate deviation of the
matrix elements from their factorized values changes the predicted BRs
by orders of magnitude. The reason is simply that some
matrix elements, which vanish in the factorization limit, 
are Cabibbo-enhanced with respect to emission
diagrams. This is precisely the mechanism which produces the large
charming-penguin enhancement~\cite{chp1,chp2}. 

\begin{figure}[t]
% GNUPLOT: LaTeX picture with Postscript
\setlength{\unitlength}{0.1bp}
\begin{picture}(3600,2160)(0,0)
\put(2008,51){\makebox(0,0){$\eta_L$}}
\put(100,1180){%
%
\makebox(0,0)[b]{\shortstack{$BR(B^+\to \rho^+K^0)\times 10^5$}}%
%
}
\put(3363,151){\makebox(0,0){1}}
\put(2821,151){\makebox(0,0){0.8}}
\put(2279,151){\makebox(0,0){0.6}}
\put(1738,151){\makebox(0,0){0.4}}
\put(1196,151){\makebox(0,0){0.2}}
\put(654,151){\makebox(0,0){0}}
\put(540,2109){\makebox(0,0)[r]{100}}
\put(540,1799){\makebox(0,0)[r]{10}}
\put(540,1490){\makebox(0,0)[r]{1}}
\put(540,1180){\makebox(0,0)[r]{0.1}}
\put(540,870){\makebox(0,0)[r]{0.01}}
\put(540,561){\makebox(0,0)[r]{0.001}}
\put(540,251){\makebox(0,0)[r]{0.0001}}
\end{picture}
\caption{\sf
Dependence of $BR(B^+\to\rho^+ K^0)$ on $\eta_L$, the
ratio of charming penguin to emission amplitudes. Exact factorization
prediction corresponds to $\eta_L=0$.
} 
\label{fig:krho}
\end{figure}
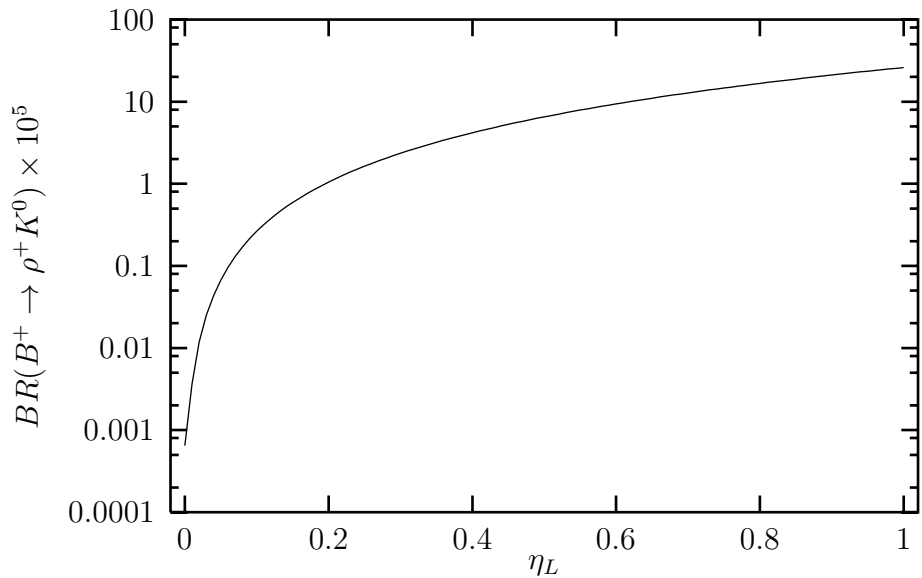

In general, it is
useful to identify two classes of operators: i) current-current
operators of the form $\bar b\Gamma q_u~\bar q_{u}^\prime\Gamma^\prime s$
which have $O(1)$ Wilson coefficients and ii) penguin operators like
$\bar b\Gamma s \sum_q \bar q\Gamma^\prime q$, the Wilson coefficients
of which are $\lappeq 0.03$ at a scale $\sim M_B$. In $B\to K\pi$ and
$B\to K\rho$, for example, the only operators
of class i) which have non-vanishing {\it factorized} matrix elements are
those containing up-quark fields. They are doubly Cabibbo-suppressed with
respect to the operators in ii), so that their Wilson coefficient enhancement
is compensated and the contribution of the two classes is comparable.
However, if we allow for a violation of the factorization, class i) operators
containing charm-quark fields can also have non-vanishing matrix elements.
Contractions of these operators, like $DP$ in fig.~\ref{fig:diag},
are called charming penguins and contribute to the decays we are considering.
Since they have large Wilson coefficients and are not Cabibbo-suppressed,
their contribution easily becomes dominant, already assuming corrections to
factorization at the level of $10-20\%$~\cite{chp1}.
Striking examples of charming-penguin dominance are given by
$B\to\rho K$ channels. For instance, $BR(B^+\to\rho^+K^0)$, shown in
fig.~\ref{fig:krho}, changes by three orders of magnitude for modest
values of $\eta_L$, which is the ratio of charming-penguin to emission
amplitudes. Exact factorization prediction, namely $\eta_L=0$, is definitely
unreliable for this decay. Notice that the value of the charming-penguin
parameters suggested by $B\to K\pi$ measurements brings the $BR(B\to
K\rho)$ near to their present experimental bounds~\cite{chp2}.

This discussion leads to the following conclusion: there is a class of
rare $B$ decays for which the use of factorized amplitudes gives very
unstable predictions, which are drastically changed by moderate corrections
to exact factorization. These decays, including
$B\to K\pi$/$K\rho$/$K\eta$, etc.
have been extensively studied in ref.~\cite{chp2}. 

The recent CLEO measurements of $BR(B\to K\pi)$~\cite{cleo} actually call for
some enhancement over the predictions obtained with factorized amplitudes.
If charming penguins have to explain the data, their value must be
about $20$-$30\%$ of the corresponding emission diagrams~\cite{chp2}. Notice
that the recent analysis of ref.~\cite{ag}, which claims to be able
to reproduce the data using factorized amplitudes, somehow takes into
account charm-loop effects in perturbation theory by enhancing
the penguin-operator Wilson coefficients. The underlying physical
process providing the enhancement is the same in the two approaches,
but we believe that the perturbative treatment is not appropriate.

A somewhat related argument, which has recently become popular, is
the effect of the final state interaction (FSI) in $B$ decays. 
In two-body decays, FSI is neglected in factorized amplitudes by definition.
Also in this case, there are amplitudes that are neglected on the
basis of factorization. For example, the annihilation diagrams, like $DA$
in fig.~\ref{fig:diag}, are usually neglected on the basis
of the following argument:
the annihilating quarks must be near enough for the weak current to
annihilate them, implying a suppression proportional to the $B$
wave function at the origin, namely a factor $f_B/M_B$.
However this and other diagrammatic arguments may not hold in presence of
FSI, which mixes up different classes of diagrams~\cite{zenc}. Since recent
theoretical estimates~\cite{dgps} suggest, at variance with factorization,
that the final-state interaction cannot be neglected in $B$ decays, many
phenomenological analyses relying on neglecting this or that amplitude on
the basis of factorization should be reconsidered. FSI effects in
$B\to \pi\pi$, $B\to K\pi$ have been studied in
the analysis of ref.~\cite{chp1}. Moreover, it has been
recently shown~\cite{matt} that rescattering effects invalidate the
Fleischer-Mannel bound~\cite{fm} on $\cos(\gamma)$ and affect bounds on new
physics~\cite{fknp}.

\subsection*{And all that}

{\small
\begin{table}[t]
\begin{center}
\begin{tabular}{|l||r|r||r|}
\hline
Channel & QCDSR & NRSX & Experiment\\
$BR\times 10^{5}$ & ($\hat\rho^2=0.65$) & ($\hat\rho^2=1.1$) & \\
\hline
$B_d\to \pi^+ D^{-}$ 	  & 301   & 331  & $310\pm 44$\\
$B_d\to \pi^+ D^{*-}$ 	  & 323   & 308  & $280\pm 41$\\
$B_d\to \rho^+ D^{-}$ 	  & 794   & 866  & $840\pm 175$\\
$B_d\to \rho^+ D^{*-}$ 	  & 994   & 949  & $730\pm 153$\\
$B^+\to \pi^+\bar D^0$ 	  & 508   & 534  & $500\pm 54$\\
$B^+\to \pi^+\bar D^{*0}$ & 605   & 567  & $520\pm 82$\\
$B^+\to \rho^+\bar D^0$   & 1015  & 112  & $1370\pm 187$\\
$B^+\to \rho^+\bar D^{*0}$& 1396  & 1339 & $1510\pm 301$\\
$B_d\to K^0 J/\Psi$ 	  & 81    & 75   & $85\pm 14$\\
$B_d\to K^{*0} J/\Psi$    & 164   & 173  & $132\pm 24$\\
$B^+\to K^+ J/\Psi$ 	  & 84    & 78   & $102\pm 11$\\
$B^+\to K^{*+} J/\Psi$    & 171   & 180  & $141\pm 33$\\
\hline
$\xi$ & 0.47 & 0.42 &\\
$\delta_\xi$ & 0.42 & 0.40 &\\
$\chi^2$/dof & 1.0 & 1.0 &\\
\hline
\end{tabular}
\end{center}
\caption[]{\sf
Predictions for measured emission-dominated decays. Two form-factor models,
NRSX and QCDSR, are considered for different
values of $\hat\rho^2$. Other required input parameters are chosen according
to the central values of ref.~\cite{chp2}. The values of the fitted
parameters $\xi$ and $\delta_\xi$ are shown in the two cases together with
the values of $\chi^2/$dof.
}
\label{tab:xi}
\end{table}
}

Charming-penguin enhancement is not always effective: there are decay
channels for which penguins are not present or not enhanced.
Table~\ref{tab:xi} contains a list of measured emission-dominated channels.
In these cases, factorized amplitudes are expected
to give more reliable predictions, yet further assumptions are required.
Using Lorentz invariance, matrix elements of currents can be parameterized in
terms of form factors, which however are known only in some special cases.
If the external states are both heavy, HQET helps to express them in terms of
few known quantities, such as the heavy masses, the Isgur-Wise function slope
$\hat\rho^2$, etc~\cite{iw}.  On the other hand, heavy-light form factors need
phenomenological models to be evaluated. Input-parameter dependence
in the heavy-heavy case, and model dependence in the heavy-light one, introduce 
further theoretical errors, which sum up with the uncertainty on factorization
and should be taken into account in the phenomenological analyses.

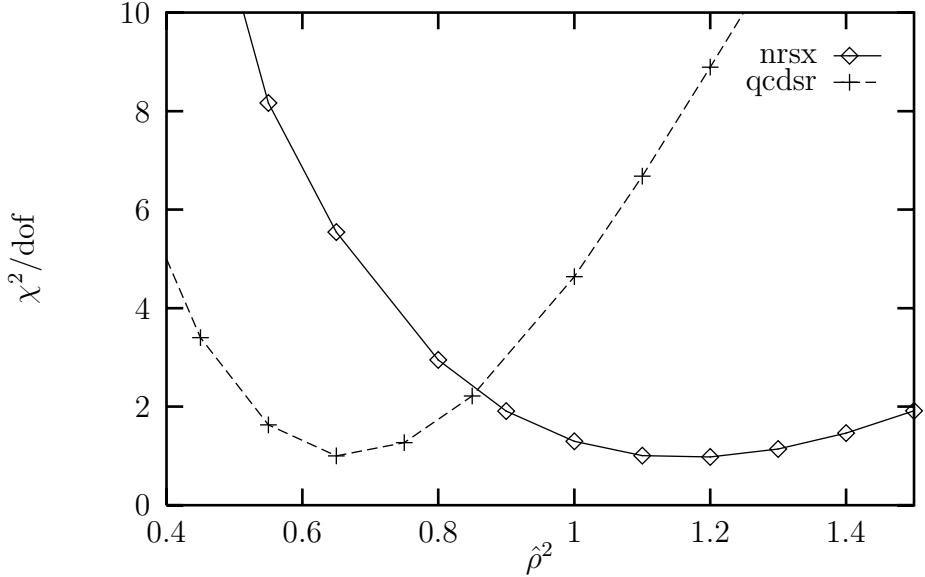
\begin{figure}[t]
% GNUPLOT: LaTeX picture with Postscript
\setlength{\unitlength}{0.1bp}
\begin{picture}(3600,2160)(0,0)
\put(3054,1846){\makebox(0,0)[r]{qcdsr}}
\put(3054,1946){\makebox(0,0)[r]{nrsx}}
\put(2008,51){\makebox(0,0){$\hat\rho^2$}}
\put(100,1180){%
%
\makebox(0,0)[b]{\shortstack{$\chi^2/\mbox{dof}$}}%
%
}
\put(3161,151){\makebox(0,0){1.4}}
\put(2649,151){\makebox(0,0){1.2}}
\put(2137,151){\makebox(0,0){1}}
\put(1624,151){\makebox(0,0){0.8}}
\put(1112,151){\makebox(0,0){0.6}}
\put(600,151){\makebox(0,0){0.4}}
\put(540,2109){\makebox(0,0)[r]{10}}
\put(540,1737){\makebox(0,0)[r]{8}}
\put(540,1366){\makebox(0,0)[r]{6}}
\put(540,994){\makebox(0,0)[r]{4}}
\put(540,623){\makebox(0,0)[r]{2}}
\put(540,251){\makebox(0,0)[r]{0}}
\end{picture}
\caption{\sf
Dependence of $\chi^2$/dof of the global fit to emission-dominated
decays on the value of the Isgur-Wise function slope $\hat\rho^2$
appearing in the heavy-heavy form factors for two different 
form-factor models.
}
\label{fig:xi2}
\end{figure}

Once form factors are chosen, no other free parameters are present
in exact factorization and emission-dominated decay rates should be predicted
without further assumptions. However it has become a standard procedure,
starting with
the well-known $a_1$ and $a_2$ of BSW~\cite{bsw}, to introduce more parameters
to account for possible deviation from factorization and fit them to the
experimental data, see e.g. refs.~\cite{chp2,cleo,ns}.
The comparison between these fitted parameters
and the factorization expectation provides a test of the reliability of
factorization. Our parameterization is the following: connected ($CE$) and
disconnected ($DE$) emissions are related according to
$CE=\xi e^{i\delta_\xi} DE$, where $\xi$ and $\delta_\xi$ are the parameters
to be fitted, then $DE$ is expressed in terms of form factors using
factorization. Colour rearrangement and exact factorization would give
$\xi=1/3$ and $\delta_\xi=0$. Fitting the decay channels in tab.~\ref{tab:xi},
we find, in agreement with other analyses~\cite{cleo,ns}, that experimental
data are
reasonably well described by factorized amplitudes for reasonable values
of $\xi$ and $\delta_\xi$. Surprisingly enough,
we have found a strong dependence on the Isgur-Wise function slope
$\hat\rho^2$ appearing in heavy-heavy form factors. Indeed, by changing
$\hat\rho^2$, the value of $\chi^2$ in the fit also changes as shown in
fig.~\ref{fig:xi2}. In the global fit, the details of the model used for the
heavy-light form
factors turn out to be hidden by this large dependence. We have considered
two popular models: NRSX~\cite{nrsx}, which is a modern version of the
original BSW model, and QCDSR~\cite{qcdsr}, a model based on light-cone
QCD sum rules. We have found
that NRSX, QCDSR and the other models considered in ref.~\cite{chp2} fit well
the data, once a
suitable value of $\hat\rho^2$ is assumed. However, for example, the best fit
within QCDSR and NRSX is given by quite different values of $\hat\rho^2$
($\hat\rho^2\sim 0.65$ and $\sim 1.1$ respectively). Moreover there exist
experimentally allowed values of $\hat\rho^2$ for which no model gives a
good fit.

It is reassuring that it is always possible to fit the data with acceptable
values of the parameters. Within a given class of two-body final states
with definite Lorentz properties, this may simply be a consequence
of the dominance of emission diagrams and of SU(3) flavour symmetry,
which reduce the number of independent amplitudes, rather than a test of the
factorization hypothesis. However the global fit of many decay channels,
including pseudoscalar and/or vector mesons in the final state, actually
probes the different form-factor models and the factorization itself.
On the other hand, the strong $\hat\rho^2$ dependence in the fit
calls for a careful study of the uncertainties affecting the factorized
amplitudes.
Indeed, the common procedure of comparing different heavy-light form-factor
models for fixed value of $\hat\rho^2$, certainly leads to underestimate the
theoretical error on $\xi$ and $\delta_\xi$ (or whatever parameters are
fitted). On the contrary, $\hat\rho^2$ should be varied within
its experimentally allowed range in order to estimate the actual
theoretical uncertainty.

\end{document}